# J-matrix method of scattering for potentials with inverse square singularity: The real representation

A. D. Alhaidari[a], H. Bahlouli[a,b], S. Al-Marzoug[a,b] and M. S. Abdelmonem[a,b]

[a] *Saudi Center for Theoretical Physics, Dhahran & Jeddah, Saudi Arabia*
[b] *Physics Department, King Fahd University of Petroleum & Minerals, Dhahran 31261, Saudi Arabia*

**Abstract**: The J-matrix method was developed to handle regular short-range scattering potentials. Its accuracy, stability, and convergence properties compare favorably with other successful scattering methods. Recently, we extended the method to the treatment of potentials with $r^{-1}$ singularity. In this work, we do the same for $r^{-2}$ singular potentials.

The study of singular potentials of $1/r^2$ type in quantum and classical mechanics is a very old subject. Classically, particles subject to such a force fall to the origin with an infinite velocity [1]. In quantum theory, however, the wavefunction oscillates indefinitely while spiraling down to the origin, leading to indeterminacy of the boundary condition at the singularity [2]. This is due to the fact that for sufficiently singular potentials the conventional methods for finding the energy eigenvalues and associated eigen-functions fail. The strong divergence of the potential at the singular point governs the leading physical behavior of the system. Consequently, the standard methods of regular quantum mechanics are supplemented by regularization or renormalization schemes – a powerful tool for studying the behavior of physical theories at various length scales [3].

Generally, the origin of the problem can be looked at from different angles and has, in fact, been tackled from different perspectives. The usual dominance of the kinetic energy term in the Hamiltonian is suppressed by the dominance of the singular interaction term near the singular point. In the case of strongly singular potentials, the wavefunction is ill behaved at the boundary leading to the lack of self-adjointness of the Hamiltonian. One way to resolve this problem is to consider the concept of self-adjoint extension of the Hamiltonian. However, such an extension is not unique and the spectrum and physical consequences depend on the chosen extension scheme (e.g., the extension parameter) [3]. A second way to resolve this problem is to use regularization and renormalization approaches. That is, the potential is regularized by introducing a short cutoff distance so that the wavefunction vanishes at this cutoff causing the removal of the undefined behavior at the singular point [4]. However, it has been found that the regularization scheme is very sensitive to the cut-off radius used to regularize the singular behavior of the potential at short distances. Physical quantities, such as scattering cross section, change dramatically with small variations in the cut-off radius and hence are not uniquely defined when this cutoff parameter is vanishingly small [5]. The issue of non-equivalence of renormalization and self-adjoint extensions for such singular interaction was also raised recently [6]. Thus, caution should be exercised while using self-adjoint extensions for the singular inverse square potential.

Aside from the above approaches there was an alternative method based on particle non-conserving processes [7]. It goes as follows: suppose we consider a particle moving in a singular potential that falls into the singularity and gets absorbed then the singular potential can be interpreted as a particle sink. This proposal is of particular importance because in a recent experiment with cold atoms moving in a $1/r^2$ dipolar potential, it was observed that some of the atoms were absorbed [8].

In this article, we isolate the inverse square singularity of the potential and transfer it to the reference Hamiltonian which already has the same behavior in the orbital term, $\ell(\ell+1)/r^2$. The virtue of the J-matrix method [9] is that its machinery could handle such singularity analytically using the theory of orthogonal polynomials; although, so far, only the orbital term was treated. In the following, we formulate the scattering theory for the $r^{-2}$ singular potential based on the J-matrix method and give an illustrative example. We restrict the treatment in this article to the case where the angular momentum quantum number $\ell$ is such that $\left(\ell+1/2\right)^2 > -A$, where $A$ is the strength of the $r^{-2}$ singularity of the potential. We call this case the real representation and leave the full fledge physical applications after the development of the complete theory with no restriction on the angular momentum (the complex representation).

The time-independent radial Schrödinger equation for a scalar point particle in the field of a central potential $V(r)$ reads as follows

$$\left[-\frac{1}{2}\frac{d^2}{dr^2}+\frac{\ell(\ell+1)}{2r^2}+V(r)-E\right]\psi_\ell(r,E)=0, \tag{1}$$

where we have used the atomic units $\hbar=m=1$. Now, we assume that the most singular part of the potential is an inverse square which can then be stripped off the potential by writing it in the form of a sum of a singular part and a regular part, $V(r)=\frac{A}{2r^2}+U(r)$, where $A$ is a real non-zero parameter and $\left|U(r)\right|<\infty$ for all $r$ [i.e. $U(r)$ is not singular[†]]. This will enable us to take the reference Hamiltonian $H_0$, in the language of J-matrix approach [9], to be $H_0=-\frac{1}{2}\frac{d^2}{dr^2}+\frac{\ell(\ell+1)+A}{2r^2}$. Therefore, the wave equation that defines the reference problem, whose analytic solution constitutes one of the main components of the J-matrix theory, is

$$\left[-\frac{1}{2}\frac{d^2}{dr^2}+\frac{\ell(\ell+1)+A}{2r^2}-E\right]\chi^\ell(r,E)=0. \tag{2}$$

If we define the wave number $k$ through the relationship $k=\sqrt{2E}$ then the solution of the above differential equation can be written as follows

$$\chi^\nu(r,E)=\sqrt{kr}\left[C_+J_{+\nu}(kr)+C_-J_{-\nu}(kr)\right], \tag{3}$$

where $J_\nu(z)$ is the Bessel function, $C_\pm$ are normalization constants and the parameter $\nu=\sqrt{\left(\ell+1/2\right)^2+A}$. Since the parameter $\nu$ is not an integer then the two independent solutions of Eq. (2) are $J_\nu(z)$ and $J_{-\nu}(z)$. However, this parameter $\nu$ can be either real or pure imaginary and in both cases these solutions are independent. In the present work, we limit ourselves to the case where $\nu$ is real; that is $\left(\ell+1/2\right)^2 > -A$. In this case, $J_{-\nu}(z)$ diverge at $z = 0$ while $J_\nu(z)$ is regular everywhere, however these solutions are asymptotically sinusoidal with the same amplitude but with a fixed phase difference. The phase difference can be made exactly $\frac{\pi}{2}$ if the irregular solution is chosen to be a particular linear combination of $J_\nu(z)$ and $J_{-\nu}(z)$ that results in the Bessel function of the second kind, $Y_\nu(z)$.

---

[†] In general, we can tolerate that $U(r)$ be 1/r singular. That is, we can allow $U(r)=\frac{B}{r}+W(r)$ with $W(r)$ non-singular and short-range. In that case, we can apply the J-matrix method with 1/r singular potentials developed recently by the founders of this approach in [16].

The present problem is a generalization of the usual J-matrix with $\ell \to \nu - \frac{1}{2}$ (non-integer). For example, the regular solution of the reference wave equation (2) is

$$\chi^\nu_{reg}(r,E) \equiv \chi^\nu_k(r) = \tfrac{2}{\sqrt{\pi}}(kr)j_{\nu-\frac{1}{2}}(kr) = \sqrt{2kr}\, J_\nu(kr), \tag{4}$$

where $j_\nu(z)$ is the spherical Bessel function. The choice of normalization is made such that $\left\langle \chi^\nu_k \middle| \chi^\nu_{k'} \right\rangle = \delta(k-k')$. Asymptotically (as $r\to\infty$) this solution is sinusoidal: $\chi^\nu_k \to \frac{2}{\sqrt{\pi}}\sin\left(kr - \pi\nu/2 + \frac{1}{4}\pi\right)$. On the other hand, for the irregular solution, where $J_\nu(z)$ in (4) changes into the Bessel function of the second kind $Y_\nu(z)$, the sine changes into the negative of a cosine asymptotically. In the context of J-matrix approach we should select a basis that will enable us to have a tridiagonal representation for the J-matrix wave operator ($J = H_0 - E$). We choose the Laguerre basis and select its parameters to ensure a tridiagonal representation. The elements of such basis are defined by

$$\phi_n(y) = A_n y^\alpha e^{-y/2} L^\beta_n(y)\,;\; y = \lambda r\,,\; n = 0,1,2,..\,, \tag{5}$$

where $A_n = \sqrt{\lambda\Gamma(n+1)/\Gamma(n+\beta+1)}$ is a normalization constant, $L^\beta_n(y)$ is the Laguerre polynomial of order $n$, $\beta > -1$, and $\lambda$ is a positive basis scale parameter. In the appendix, we show how our present approach can also be formulated in another equivalent basis; the oscillator basis. Using the differential equation of the Laguerre polynomials [10] and their differential formula, $y\frac{d}{dy}L^\lambda_n = n\,L^\lambda_n - (n+\lambda)L^\lambda_{n-1}$, we can write

$$\begin{aligned}\left(H_0 - E\right)\left|\phi_n\right\rangle = \frac{\lambda^2}{2}\left[\frac{n}{y}\left(1+\tfrac{\beta+1-2\alpha}{y}\right)+\frac{\alpha}{y}-\frac{1}{4}+\frac{\ell(\ell+1)+A-\alpha(\alpha-1)}{y^2}-\frac{2E}{\lambda^2}\right]\left|\phi_n\right\rangle \\ +\frac{n+\beta}{y}\left(\tfrac{2\alpha-\beta-1}{y}\right)\frac{A_n}{A_{n-1}}\left|\phi_{n-1}\right\rangle\end{aligned} \tag{6}$$

Imposing the tridiagonal representation constraint on the operator $J = H_0 - E$ requires that $2\alpha = \beta + 1$ and $\beta = 2\nu$. Using the recursion relation of the Laguerre polynomials and their orthogonality property, we obtain the following tridiagonal representation of the reference wave operator

$$\begin{aligned}\frac{8}{\lambda^2}J^\nu_{nm}(E) = \frac{8}{\lambda^2}\left\langle\phi^\nu_n\right|\left(H_0 - E\right)\left|\phi^\nu_m\right\rangle = (2n+2\nu+1)\left(1-4\mu^2\right)\delta_{n,m} + \\ \left(1+4\mu^2\right)\left[\sqrt{n(n+2\nu)}\delta_{n,m+1}+\sqrt{(n+1)(n+2\nu+1)}\,\delta_{n,m-1}\right]\end{aligned} \tag{7}$$

where $\mu = k/\lambda$. Expanding the regular solution, also called sine-like solution in the J-matrix formalism, in the above basis gives

$$\chi_{\sin}^{\nu}(r,E) \equiv \chi_{reg}^{\nu}(r,E) = \sum_{n=0}^{\infty} s_n(E)\phi_n^{\nu}(y), \qquad (8)$$

We can easily show that the tridiagonal nature of the J-matrix operator (7) results in the following three-term recursion relation for the expansion coefficients

$$\left(2n+2\nu+1\right)\cos\theta\; s_n = \sqrt{n(n+2\nu)}\; s_{n-1} + \sqrt{(n+1)(n+2\nu+1)}\; s_{n+1}, \qquad (9a)$$

with the following initial condition

$$\left(2\nu+1\right)\cos\theta\; s_0 = \sqrt{2\nu+1}\; s_1 \qquad (9b)$$

where $\cos\theta = \frac{4\mu^2-1}{4\mu^2+1}$. Multiplying both sides of the above recursion relation by $\sqrt{\Gamma(n+2\nu+1)/\Gamma(n+1)}$ transforms it into that for the classical Gegenbauer (ultra-spherical) polynomial $C_n^{\nu+\frac{1}{2}}(x)$, where $x = \cos\theta$. This implies that $s_n(E)$ is equal to $\sqrt{\Gamma(n+1)/\Gamma(n+2\nu+1)}\, C_n^{\nu+\frac{1}{2}}(x)$ multiplied by an arbitrary non-zero function of the energy that is independent of *n*.

At this point we divert from the traditional approach in the J-matrix method which uses the above recursion relation to derive the second order differential equation obeyed by $\{s_n\}$. Instead, we evaluate $\{s_n\}$ using the orthogonality property of the Laguerre polynomials and direct integration of (8) after multiplying both sides by $y^{-1}\phi_n(y)$ giving

$$s_n(E) = \frac{A_n}{\lambda}\int_0^{\infty} y^{\nu-\frac{1}{2}} e^{-y/2} L_n^{2\nu}(y)\,\chi_{reg}^{\nu}(y/\lambda, E)\,dy\,, \qquad (10)$$

Rewriting the wavefunction in terms of the Bessel function as expressed in (4) and substituting in (10) we obtain

$$s_n(E) = \sqrt{2\mu}\,\frac{A_n}{\lambda}\int_0^{\infty} y^{\nu} e^{-y/2} L_n^{2\nu}(y)\, J_{\nu}(\mu y)\,dy\,, \qquad (11)$$

This integral is not found in mathematical tables but has been evaluated by one of the authors in [11]. The result is

$$s_n(E) = \Gamma\left(\nu+\tfrac{1}{2}\right)\frac{A_n}{\lambda\sqrt{\pi}}(2\sin\theta)^{\nu+\frac{1}{2}} C_n^{\nu+\frac{1}{2}}(\cos\theta), \qquad (12)$$

Using the recursion properties of the Gegenbauer polynomials one can easily verify that these expansion coefficients obey the recursion relation (9a) and its associated initial condition (9b). The above solution of the three-term recursion relation is called the sine-like solution and obeys the homogeneous initial condition (9b). In order to obtain the second independent solution of the recursion relation (9a), called the cosine-like solution, we impose that its expansion coefficients meet the following criteria:

1. Satisfy the same recursion relation (9a) as the sine-like coefficients except for the initial relation (9b).
2. The initial relation (9b) is chosen to make the asymptotic behavior of the cosine-like solution identical to that of the irregular solution (i.e., sinusoidal with the same amplitude as the sine-like but with a $\pi/2$ phase difference).

This second independent solution is a regularized (at the origin) version of the irregular solution and it solves the original differential equation (2) only asymptotically. We write it as $\chi_{\cos}(r,E)=\sum_{n=0}^{\infty}c_n(E)\phi_n(r)$ where the expansion coefficient obey the following initial inhomogeneous relation

$$J_{00}\,c_0+J_{01}\,c_1=\gamma\,, \tag{13}$$

where $\gamma$ is an energy dependent function which is obtained by imposing the asymptotic boundary condition (second criterion above) that results in $\gamma=-W/2s_0$ with $W$ being the Wronskian of the two independent solutions of Eq. (2). Using these two solutions, we obtain $W=4k/\pi=\frac{2\lambda}{\pi}\sqrt{\frac{1+x}{1-x}}$ while $s_0=2^{\nu+\frac{1}{2}}\frac{\Gamma(\nu+1/2)}{\sqrt{\lambda\pi\Gamma(2\nu+1)}}\left(1-x^2\right)^{\frac{\nu}{2}+\frac{1}{4}}$. Moreover, $c_n$ obey the recursion relation (9a) which can be brought into a simpler form through the transformation $c_n(x)=\sqrt{\Gamma(n+1)/\Gamma(n+2\nu+1)}\,G_n(x)$ leading to

$$\left(2n+2\nu+1\right)x\,G_n=(n+2\nu)\,G_{n-1}+(n+1)\,G_{n+1}\quad;\quad n=1,2,3,.... \tag{14a}$$

The initial condition (13) can then be written explicitely as follows

$$-(2\nu+1)xG_0+G_1=\frac{2\gamma\sqrt{\Gamma(2\nu+1)}}{\lambda^2\left(\mu^2+1/4\right)}. \tag{14b}$$

Iterating the above recursion relation starting from the initial condition (14b) for successive values of $n$ leads to the following general structure of the solution

$$G_n(x)=G_0(x)\,C_n^{\nu+1/2}(x)+\eta(x)\,Q_{n-1}(x)\,, \tag{15}$$

where $\eta=\frac{2\gamma\sqrt{\Gamma(2\nu+1)}}{\lambda^2\left(\mu^2+1/4\right)}=\frac{-2^{\frac{3}{2}-\nu}}{\sqrt{\lambda\pi}}\frac{\Gamma(2\nu+1)}{\Gamma\left(\nu+1/2\right)}(1-x^2)^{\frac{1}{4}-\frac{\nu}{2}}$ and the new polynomials $Q_n$ satisfy the following recursion relation

$$\begin{aligned}&\left(2n+2\nu+3\right)x\,Q_n=(2\nu+n+1)Q_{n-1}+(n+2)Q_{n+1};\quad n\geq 1\\ &\left(2\nu+3\right)xQ_0-2Q_1=0\quad;\quad Q_0=1\quad;\ Q_{-1}\equiv 0\end{aligned} \tag{16}$$

Comparing this recursion relation for $Q_n$ with that associated with the Gegenbauer polynomial $C_n^{\nu+1/2}$, we deduce that their recursion coefficients are shifted by one unit (i.e. $n\to n+1$). These are just the associated Gegenbauer or Wimp polynomials defined in Sec. 5.7.1 of [12] as $\mathcal{C}_n^{\nu+1/2}(x,1)$. In the literature, $\mathcal{C}_n^{\nu}(x,c)$ obeys the following recursion relation

$$\begin{aligned}&2\left(n+\nu+c\right)x\,\mathcal{C}_n^{\nu}=\left(2\nu+n+c-1\right)\mathcal{C}_{n-1}^{\nu}+\left(n+c+1\right)\mathcal{C}_{n+1}^{\nu};\quad n\geq 0\\ &\mathcal{C}_0^{\nu}(x,c)=1\,;\quad \mathcal{C}_{-1}^{\nu}(x,c)=0.\end{aligned} \tag{17}$$

Thus, the general solution for the recursion relation (14) can be written as follows

$$G_n(x)=G_0(x)C_n^{\nu+1/2}\left(x\right)+\eta(x)\,\mathcal{C}_{n-1}^{\nu+1/2}\left(x,1\right). \tag{18}$$

Hence, the cosine-like coefficients become

$$c_n(E)=\tau(x)s_n(E)+\eta(x)\sqrt{\frac{\Gamma(n+1)}{\Gamma(n+2\nu+1)}}\,\mathcal{C}_{n-1}^{\nu+1/2}\left(x,1\right), \tag{19}$$

where we have rewritten $G_0$ as $G_0(x)=\tau(x)\sqrt{\Gamma(2\nu+1)}\,s_0(x)$ with $\tau(x)$ being an arbitrary real function of the energy to be determined below. It should be obvious that for $\gamma$ = 0 these coefficients, which now become $c_n(E)=\tau(x)s_n(E)$, satisfy the homogenous recursion relation (9) for any $\tau$.

Using the differential equation of the Gegenbauer polynomials, one can show that the sine-like coefficients (12) satisfy the following second order energy differential equation

$$\hat{D}s_n(E) = \left[\left(1-x^2\right)\frac{d^2}{dx^2} - x\frac{d}{dx} - \frac{\nu^2 - 1/4}{1-x^2} + \left(n+\nu+\tfrac{1}{2}\right)^2\right]s_n(E) = 0. \qquad (20)$$

The cosine-like coefficients $c_n(E)$ should also satisfy the same differential equation (i.e., $\hat{D}c_n = 0$). Applying the operator $\hat{D}$ on (19) for $n$ = 0 gives $\hat{D}\left(\tau s_0\right) = 0$, which upon the use of $\hat{D}s_0 = 0$ results in the following differential equation for $\tau$

$$\left[(1-x^2)\frac{d}{dx} + 2(1-x^2)\frac{s_0'}{s_0} - x\right]\frac{d\tau}{dx} = 0, \qquad (21)$$

where $s_0' = ds_0/dx$ . Using the explicit from of $s_0(x)$ this equation becomes

$$\left[(1-x^2)\frac{d}{dx} - 2(\nu+1)x\right]\frac{d\tau}{dx} = 0, \qquad (22)$$

whose solution is $d\tau/dx = R_\nu(1-x^2)^{-\nu-1}$, where $R_\nu$ is a constant that depends only on $\nu$. Integration gives

$$\tau(x) = x R_\nu \, {}_2F_1\left(\tfrac{1}{2}, \nu+1; \tfrac{3}{2}; x^2\right), \qquad (23)$$

where ${}_2F_1\left(a,b;c;z\right)$ is the hypergeometric function. The constant $R_\nu$ is obtained by matching the irregular solution to the cosine-like solution asymptotically giving $R_\nu = 2\Gamma(\nu+1)/\sqrt{\pi}\Gamma\left(\nu+1/2\right)$. Finally, with all elements defined we can write the cosine-like solution as

$$\begin{aligned}\chi_{\cos}(r,x) &= \sum_{n=0}^{\infty} c_n(x)\phi_n(r) \\ &= \tau(x)\chi_{\sin}(r,E) + \eta(x)\sum_{n=0}^{\infty}\sqrt{\frac{\Gamma(n+1)}{\Gamma(n+2\nu+1)}}\, C_{n-1}^{\nu+1/2}\left(x,1\right)\phi_n(\lambda r)\end{aligned} \qquad (24)$$

To verify the validity of our approach we perform computations of the sine-like and cosine-like solutions defined by their respective expansions

$$\chi_{\sin}(r,E) = \sum_{n=0}^{\infty} s_n(E)\phi_n(r), \quad \chi_{\cos}(r,E) = \sum_{n=0}^{\infty} c_n(E)\phi_n(r). \qquad (25)$$

Both should be regular at the origin and $\chi_{\sin}(r,E) = \chi_{reg}^{\nu}(r,E) = \sqrt{2kr}\, J_\nu(kr)$ whereas $\lim_{r\to\infty}\chi_{\cos}(r,E) = \lim_{r\to\infty}\chi_{irr}^{\nu}(r,E) = \lim_{r\to\infty}\sqrt{2kr}\, Y_\nu(kr)$. For illustration, we show in Figure 1 the sine-like and cosine-like solutions obtained from the series expansions (25) for a given energy, angular momentum and singularity strength $A$. As expected, the sine-like solution is identical to the regular solution (4) everywhere while the cosine-like solution coincides with the irregular solution only asymptotically but deforms itself near the origin, as imposed by the J-matrix regularization scheme, to comply with the vanishing of the wavefunction at this boundary. On the other hand, Figure 2 shows that the phase difference between these two solutions is exactly $\pi/2$ asymptotically as required by construction.

To illustrate our findings, we select the following example for the short-range potential

$$U(r) = 7.5 r^2 e^{-r}, \tag{26}$$

which has been studied frequently in the literature [13-14]. For ease of comparison with results in the literature, we have selected the inverse square potential strength $A$ an integer such that $\ell(\ell+1)+A = L(L+1)$ where $L$ is also an integer. As shown in Table 1 our results are in good agreement with those obtained in [14] for $L$= 0,1,2 despite the fact that we have used a basis size of 100 whereas the size of the basis in [14] is 200. On the other hand, the resonances in reference [14] were generated using the finite complex rotation method and the matrix elements of the scattering potential (26) were evaluated using the Gauss quadrature approach. However, in our scheme we used the J-matrix approach to evaluate the resonances as being the poles of corresponding S-matrix. In fact, in order to find resonance and bound state energies we use the method based on the J-matrix calculation of the scattering S-matrix, $S(E)$, in the complex energy plane which is given by [15,16]

$$S(E) = T_{N-1}(E)\,\frac{1+g_{N-1,N-1}(E)\,J_{N-1,N}(E)R_N^-(E)}{1+g_{N-1,N-1}(E)\,J_{N-1,N}(E)R_N^+(E)}, \tag{27}$$

where the J-matrix kinematics quantities $\{T_n, R_n^\pm\}$, the finite Green's function $g_{N-1,N-1}$ and $J_{N-1,N}$ are given in Table 2 for the Laguerre basis. Bound states and resonances energies are then defined as the roots of equation $S^{-1}(E) = 0$.

The stability and accuracy of the theoretical scheme that we used are mainly dependent on two parameters, the length scale parameter $\lambda$ and the basis space dimension, $N$. It is important to choose the proper range of values of $\lambda$ where the bound and resonant energies are stable (i.e., independent of the value of $\lambda$ in this range) [17].

In summary, we have investigated the inverse square potential using the J-matrix machinery. Basically we have combined the inverse square singularity with the orbital term, $\ell(\ell+1)/r^2$, which already exists in the reference Hamiltonian. The virtue of the J-matrix method is that it could handle such singularity analytically using the theory of orthogonal polynomials as shown above. To test the validity of our approach we have computed the resonance energies associated with the potential $V(r)$ whose regular part is given by (26) and compared our results with those in the literature. We have limited our investigation in this work to real representations where the strength of the inverse square potential is such that $\ell(\ell+1)+A>0$; otherwise we will end up with Bessel functions of imaginary order. Such general treatment requires an involved approach and deserves future efforts.

**Acknowledgements:**

We acknowledge the support of King Fahd University of Petroleum and Minerals under group project number RG1109-1 and RG1109-2. Additional support by The Saudi Center for Theoretical Physics (SCTP) is highly appreciated by all authors. We also appreciate Dr. I. M. Nasser help in checking the results shown in Table 1.

**Appendix: Solution of the reference problem in the oscillator basis**

In this appendix, we evaluate the solutions of the reference Hamiltonian, $H_0$, using the oscillator basis defined by

$$\phi_n(y) = A_n y^{\alpha} e^{-y/2} L_n^{\beta}(y); \; y = (\lambda r)^2, \tag{A1}$$

where $\alpha$ is a positive parameter, $\beta > -1$, and the normalization constant $A_n$ is chosen in this basis as $A_n = \sqrt{2\lambda\Gamma(n+1)/\Gamma(n+1+\beta)}$. The reference Hamiltonian $H_0$ is just the free kinetic energy operator, $H_0 = -\frac{1}{2}\frac{d^2}{dr^2} + \frac{\ell(\ell+1)+A}{2r^2}$. Therefore, the reference wave equation for the J-matrix is defined by

$$\begin{aligned} J\chi(r,E) &= \left[-\frac{1}{2}\frac{d^2}{dr^2} + \frac{\ell(\ell+1)+A}{2r^2} - E\right]\chi(r,E), \\ J\chi(y,\mu) &= \lambda^2\left[-2y\frac{d^2}{dy^2} - \frac{d}{dy} + \frac{\ell(\ell+1)+A}{2y} - \frac{\mu^2}{2}\right]\chi(y,\mu) = 0, \end{aligned} \tag{A2}$$

where $\mu = k/\lambda = \sqrt{2E}/\lambda$ and $J = H_0 - E$. Using the differential equation of the Laguerre polynomials, their differential formula and recursion relation [10], we can show that the action of the wave operator $J$ will be given by

$$\begin{aligned} \frac{1}{\lambda^2}J(\mu)|\phi_n\rangle = \Bigg[&\frac{-(\alpha+n)+\left[\ell(\ell+1)+A\right]/2+2n(\beta+1-2\alpha)-2\alpha(\alpha-1)}{y} \\ &+\frac{1}{2}+2n+2\alpha-\frac{y}{2}-\frac{\mu^2}{2}\Bigg]|\phi_n\rangle + \frac{2(n+\beta)}{y}\frac{A_n}{A_{n-1}}\left(2\alpha-\beta-\tfrac{1}{2}\right)|\phi_{n-1}\rangle \end{aligned} \tag{A3}$$

Such a matrix representation for $J$ will be tridiagonal only if $\beta = 2\alpha - \frac{1}{2}$ and $\beta^2 = \left(\ell+\frac{1}{2}\right)^2 + A$, which then results in

$$\begin{aligned} &\frac{2}{\lambda^2}J_{nm}(\mu) = \frac{2}{\lambda^2}\langle\phi_n|H_0 - E|\phi_m\rangle = \\ &\left[\left(2n+\nu+1-\mu^2\right)\delta_{n,m} + \sqrt{n(n+\nu)}\,\delta_{n,m+1} + \sqrt{(n+1)(n+\nu+1)}\,\delta_{n,m-1}\right] \end{aligned} \tag{A4}$$

with $\nu = \beta = \sqrt{\left(\ell+\frac{1}{2}\right)^2 + A}$. We can identify the J-matrix elements from the above relation as follows

$$J_{n,n} = \frac{\lambda^2}{2}\left(2n+\nu+1-\mu^2\right), \; J_{n,n+1} = J_{n+1,n} = \frac{\lambda^2}{2}\sqrt{(n+1)(n+\nu+1)}\,. \tag{A5}$$

Writing the regular solution as $\chi^{\nu}_{reg}(r,E) \equiv \chi^{\nu}_{\sin}(r,E) = \sum_{n=0}^{\infty} s_n(E)\phi_n(y)$ leads to the following recursion relation for the expansion coefficients

$$\left(2n+\nu+1-\mu^2\right)s_n + \sqrt{n(n+\nu)}s_{n-1} + \sqrt{(n+1)(n+\nu+1)}s_{n+1} = 0\,, \tag{A6a}$$

$$\left(\nu+1-\mu^2\right)s_0 + \sqrt{\nu+1}\,s_1 = 0\,. \tag{A6b}$$

We use the following transformation to get rid of the square roots in the above recursion relation

$$s_n(E) = B_\nu(-1)^n\sqrt{\frac{\Gamma(n+1)}{\Gamma(n+\nu+1)}}L_n^{\nu}(x)\,, \tag{A7}$$

where $x=\mu^2$ and $B_\nu$ is a normalization constant that dependents on $\nu$ and will be fixed later. This transformation maps (A6) into the following recursion relation for the Laguerre polynomials

$$\left(2n+\nu+1-x\right)L_n^\nu-(n+\nu)L_{n-1}^\nu-(n+1)L_{n+1}^\nu=0\,. \tag{A8}$$

In analogy to what we did in the main text for the Laguerre basis, we use direct integration and identify the regular solution of the reference wave equation with the sine-like solution

$$\begin{aligned}\chi_{reg}(y,\mu)&=\sqrt{2\mu}y^{1/4}J_\nu(\mu\sqrt{y})=\sum_{n=0}^{\infty}s_n(\mu)\phi_n(y)\\&=\sum_{n=0}^{\infty}A_ns_n(\mu)y^{(\nu+\frac{1}{2})/2}e^{-y/2}L_n^\nu(y).\end{aligned} \tag{A9}$$

Using the orthogonality of the Laguerre polynomials, we obtain

$$s_n(E)=\sqrt{\frac{\mu\Gamma(n+1)}{\lambda\Gamma(n+\nu+1)}}\int_0^\infty y^{\frac{\nu}{2}}e^{-y/2}J_\nu(\mu\sqrt{y})L_n^\nu(y)dy\,. \tag{A10}$$

The result of this integral can be obtained from [18] to give

$$s_n(E)=2(-1)^n\sqrt{\frac{\Gamma(n+1)}{\lambda\Gamma(n+\nu+1)}}\,\mu^{\nu+\frac{1}{2}}e^{-\mu^2/2}L_n^\nu(\mu^2)\,. \tag{A11}$$

From this exact result, we can deduce the pre-factor in the transformation (A7) as $B_\nu=\frac{2}{\sqrt{\lambda}}\mu^{\nu+\frac{1}{2}}e^{-\mu^2/2}$. The expansion coefficients of the second regularized cosine-like solution, $\chi_{\cos}^\nu(r,E)\equiv\sum_{n=0}^{\infty}c_n(E)\phi_n(y)$, obey the same recursion relation (A6a) but with the following inhomogeneous initial condition

$$J_{00}c_0+J_{01}c_1=\gamma\quad;\quad\gamma=-\frac{W}{2s_0}=-\frac{\lambda}{\pi}\sqrt{\lambda\Gamma(\nu+1)}\mu^{-\nu+\frac{1}{2}}e^{\mu^2/2}\,, \tag{A12}$$

where $W$ is the Wronskian of the two independent solutions. Proceeding in the same way as we did for the Laguerre basis in the main text, we can write the general solution of the recursion relation (A6a) with the initial relation (A12) in the following form

$$c_n(E)=\tau(x)s_n(E)+\eta(x)P_{n-1}(x)\,, \tag{A13}$$

where $\tau(x)$ is arbitrary and the new polynomials $P_n(x)$ obey the homogeneous recursion relation (A6) but with the recursion coefficients shifted by one unit (i.e., $n\to n+1$) and $P_{-1}\equiv 0$. It should be obvious that for $\gamma$ = 0 the coefficients (A13) satisfy the homogenous recursion relation (A6) for any $\tau$. Writing these polynomials as

$$P_n(x)=(-1)^n\sqrt{\frac{\Gamma(n+2)}{\Gamma(n+\nu+2)}}L_n^\nu(x,1)\,, \tag{A14}$$

with $L_0^\nu(x,1)=1$, shows that $L_n^\nu(x,1)$ obey the same recursion relation as the Laguerre polynomials (A8) but with a recursion coefficient shifted by one unit. These are the Laguerre polynomials of the second kind; also known as the Wimp polynomials [12]. Substituting (A13) in the initial relation (A12) gives $\eta=\gamma/J_{01}P_0=-\frac{2/\pi}{\sqrt{\lambda}}\Gamma(\nu+1)\mu^{-\nu+\frac{1}{2}}e^{\mu^2/2}$. Thus, to determine the cosine-like expansion coefficient completely we only need to calculate $\tau(x)$. Now, using the differential equation of the Laguerre polynomials, one can show that the sine-like coefficients (A11) satisfy the following second order energy differential equation

$$\hat{D}\, s_n(E) = \left[ x\,\frac{d^2}{dx^2} + \frac{1}{2}\frac{d}{dx} - \frac{\nu - 1/4}{4x} - \frac{1}{4}x + \frac{1}{2}\left(2n+\nu+1\right)\right] s_n(E) = 0\,. \qquad \text{(A15)}$$

The cosine-like coefficients $c_n(E)$ should also satisfy the same differential equation (i.e., $\hat{D}\, c_n = 0$). Applying the operator $\hat{D}$ on (A13) for $n$ = 0 gives $\hat{D}\left(\tau s_0\right) = 0$, which upon the use of $\hat{D}\, s_0 = 0$ results in the following differential equation for $\tau$

$$\left[ x\,\frac{d}{dx} + 2x\,\frac{s_0'}{s_0} + \frac{1}{2}\right]\frac{d\tau}{dx} = 0\,, \qquad \text{(A16)}$$

Using the explicit from of $s_0(x)$ from (A11), this equation becomes

$$\left[ x\,\frac{d}{dx} - x + \left(\nu + \tfrac{1}{2}\right)\right]\frac{d\tau}{dx} = 0\,, \qquad \text{(A17)}$$

whose solution is $d\tau/dx = R_\nu \mu^{-(2\nu+1)} e^{\mu^2}$, where $R_\nu$ is a constant. Integrating, we obtain

$$\begin{aligned} \tau(E) &= R_\nu\,(-1)^{\nu+\frac{1}{2}}\,\Gamma\left(-\nu+\tfrac{1}{2}, -\mu^2\right) \\ &= \frac{R_\nu}{\nu - 1/2}\,\mu^{-2\nu+1}\,{}_1F_1\left(-\nu+\tfrac{1}{2}; -\nu+\tfrac{3}{2}; \mu^2\right) \end{aligned} \qquad \text{(A18)}$$

where $\Gamma(\alpha, z)$ is the incomplete gamma function and ${}_1F_1\left(a;c;z\right)$ is the confluent hypergeometric function. Finally, we can write

$$\begin{aligned} \chi_{\cos}(r,x) &= \sum_{n=0}^{\infty} c_n(x)\,\phi_n(r) \\ &= \tau(E)\,\chi_{\sin}(r,E) + \eta(E)\sum_{n=0}^{\infty}(-1)^n\sqrt{\frac{\Gamma(n+1)}{\Gamma(n+\nu+1)}}\,L_{n-1}^{\nu}(\mu^2,1)\,\phi_n(\lambda r) \end{aligned} \qquad \text{(A19)}$$

Now, with all sine- and cosine-like coefficients being determined, the analytic part of the J-matrix method in the Oscillator basis is completely defined. The finite numerical part is obtained by the standard J-matrix tools of using the Gauss quadrature to evaluate the matrix elements of the non-singular short-range potential component, *U*(*r*), in the Oscillator basis.

**Table Captions:**

**Table 1**: Resonance energies for the potential $V(r)$ whose regular part is given by Eq. (26). Our results are compared with those found in reference [14]. The accuracy of our work is relative to a basis dimension of $N = 100$, whereas in [14] $N = 200$.

**Table 2:** The J-matrix kinematics quantities $\left\{T_n(E), R_n^{\pm}(E)\right\}$, the finite Green's function $g_{N-1,N-1}(z)$, and the reference wave operator matrix element $J_{N-1,N}(E)$ all in the Laguerre basis. The energy dependent quantity $\eta$ is defined in the text below Eq. (15). $\left\{\varepsilon_n\right\}_{n=0}^{N-1}$ is the set of eigenvalues of the finite $N \times N$ total Hamiltonian $H$ and $\left\{\tilde{\varepsilon}_n\right\}_{n=0}^{N-2}$ are the eigenvalues of the truncated $H$ obtained by deleting the last row and last column.

**Figures Captions:**

**Fig. 1**: Plot of the sine-like, cosine-like, regular, and irregular solutions for $\ell = 1$, $A = 2.0$, and $\mu = 1.5$. The asymptotic amplitude of these sinusoidal solutions is $2/\sqrt{\pi}$.

**Fig. 2**: Asymptotic plot of the sine-like and cosine-like solutions for the same physical parameters as in Fig. 1.

**Table 1**

| $\{\ell, A, L\}$ | This work | Ref. [14] |
|---|---|---|
| {1,−2,0} | 5.064929607 −5.976034787i | 5.064929608 −5.976034788i |
| | 4.268860299 −8.716908433i | 4.268860299 −8.716908434i |
| | 2.9477816003 −11.530514731i | 2.9477816003 −11.530514731i |
| | 1.147183738 −14.369007137i | 1.147183738 −14.369007137i |
| | −1.096687 −17.2010213i | −1.096688979 −17.2010217935i |
| {2, −4,1} | 5.4277422972 −4.640344486i | 5.4277422973 −4.640344487i |
| | 5.3604696511 −2.197241165i | 5.3604696511 −2.197241165 i |
| | 4.8877690564 −7.3118076603i | 4.8877690564 −7.3118076605 i |
| | 4.6466344207 −0.3252945142i | 4.6466344207 −0.3252945143i |
| | 3.8017984630 −10.0935143203i | 3.8017984630 −10.0935143205i |
| {1,4,2} | 5.7936930647 −3.3304758659i | 5.7936930648 −3.330475866 i |
| | 5.5029439382 −5.9215612776i | 5.5029439380 −5.9215612775i |
| | 5.4913453119 −1.05364578685i | 5.4913453119 −1.05364578685i |
| | 4.6527742297 −8.6624552084i | 4.6527742298 −8.6624552085i |
| | 3.2892316412 −11.474735954i | 3.2892316413 −11.474735955i |

**Table 2**

| | |
|---|---|
| $g_{N-1,N-1}(z)$ | $\frac{1}{N+2\nu}\left[\prod_{m=0}^{N-2}(\tilde{\varepsilon}_m - z)\Big/\prod_{n=0}^{N-1}(\varepsilon_n - z)\right]$ |
| $T_n(E)$ | $\frac{c_n(E)-\mathrm{i}s_n(E)}{c_n(E)+\mathrm{i}s_n(E)}$ |
| $R_n^{\pm}(E)$ | $\frac{c_n(E)\pm\mathrm{i}s_n(E)}{c_{n-1}(E)\pm\mathrm{i}s_{n-1}(E)}$ |
| $s_n(E)$ | $\Gamma\left(\nu+\frac{1}{2}\right)\frac{A_n}{\lambda\sqrt{\pi}}(2\sin\theta)^{\nu+\frac{1}{2}}C_n^{\nu+\frac{1}{2}}(\cos\theta)$ |
| $c_n(E)$ | $s_n(E)-\eta\sqrt{\frac{\Gamma(n+1)}{\Gamma(n+2\nu+1)}}\,\mathcal{C}_{n-1}^{\nu+1/2}\left(\cos\theta,1\right)$ |
| $J_{N-1,N}(E)$ | $\left(E+\lambda^2/8\right)\sqrt{N(N+2\nu)}$ |

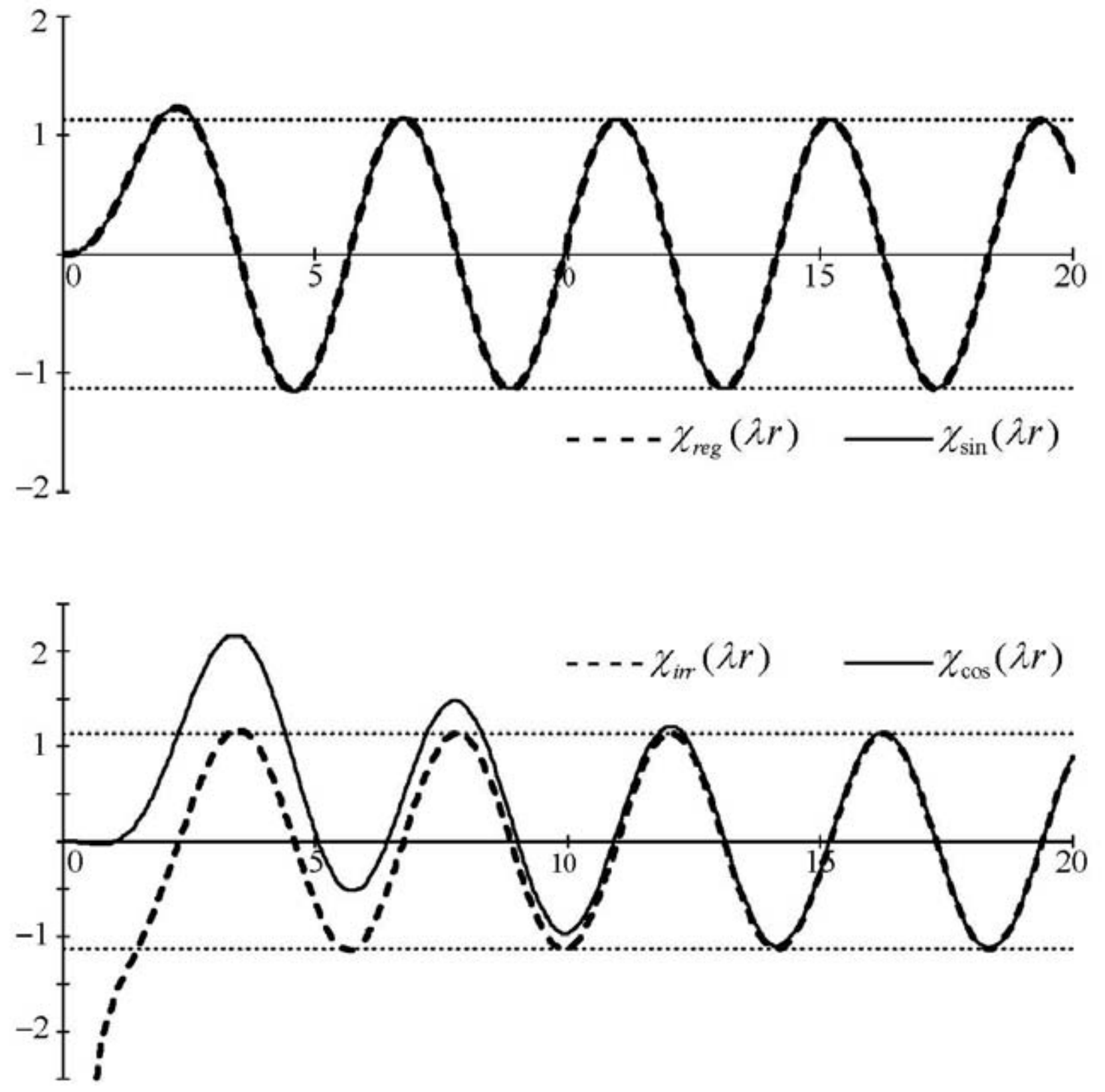

**Fig. 1**

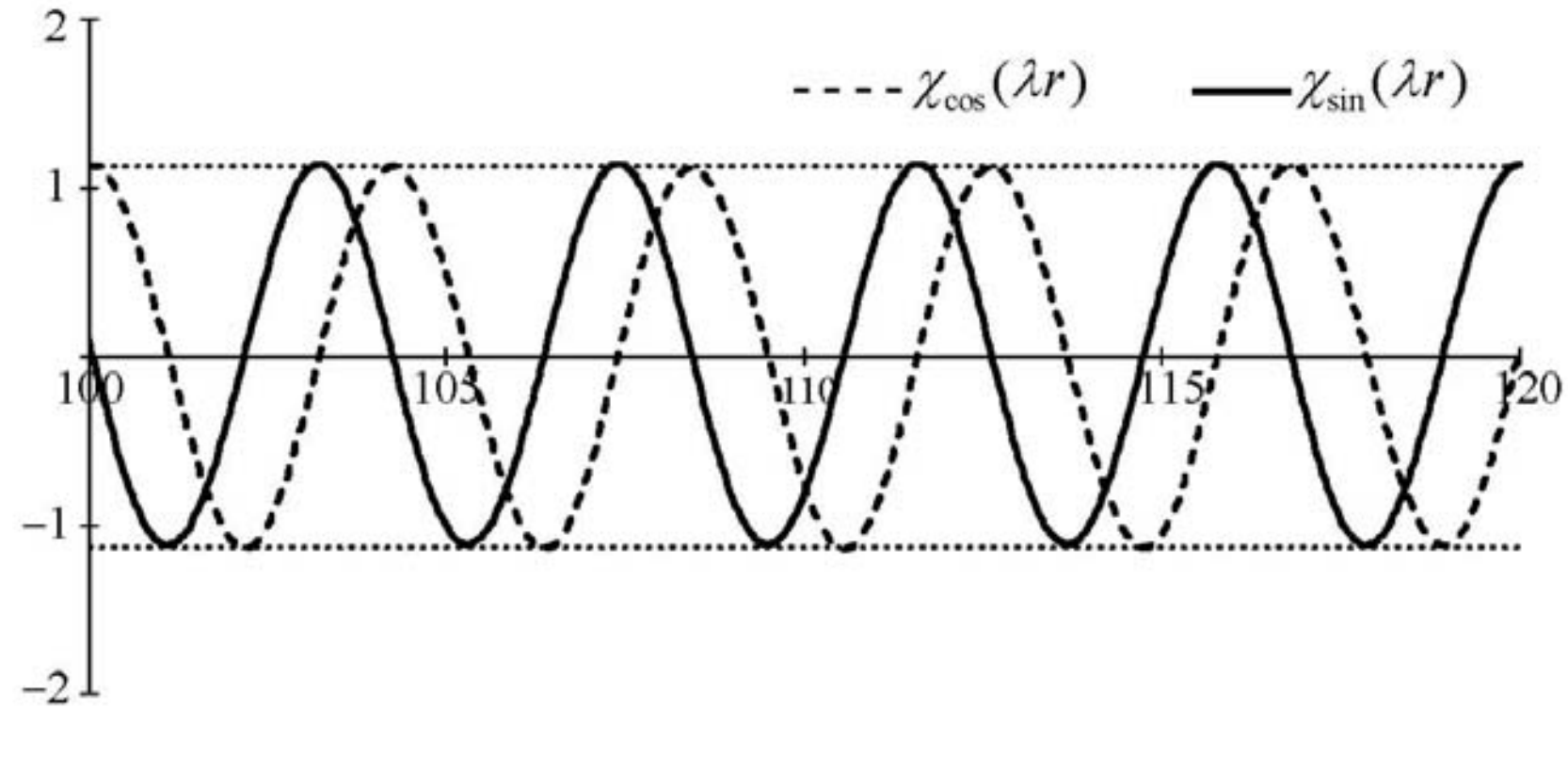

**Fig. 2**